\documentclass{./styles/svproc}
\usepackage{url}
\usepackage{graphicx}
\usepackage{float}

\usepackage[margin=1in]{geometry}
\usepackage{amsmath}
\usepackage{amssymb}
\usepackage{booktabs}
\usepackage{multirow}
\usepackage{algorithm}
\usepackage{algorithmic}
\usepackage{authblk}
\usepackage{subcaption}

\begin{document}
\mainmatter              

\title{Collective Intelligence with Foundation Models}

\titlerunning{Collective Intelligence with Foundation Models}

\author{J. de Curt\`o\inst{1,2} \and I. de Zarz\`a\inst{3}}

\authorrunning{de Curt\`o et al.}

\tocauthor{de Curt\`o et al.}

\institute{
Department of Computer Applications in Science \& Engineering, BARCELONA Supercomputing Center, 08034 Barcelona, Spain 
\and 
Escuela Técnica Superior de Ingeniería (ICAI), Universidad Pontificia Comillas, 28015 Madrid, Spain
\and Human Centered AI, Data \& Software, LUXEMBOURG Institute of Science and Technology, L-4362 Esch-sur-Alzette, Luxembourg}

\maketitle              

\begin{abstract}

As foundation models grow in scale and diversity, coordinating multiple models into cooperative reasoning systems offers a promising path toward safer and more globally reliable AI. This chapter presents a multi-agent reasoning framework where multiple solver models generate independent drafts, each undergoes structured critique and revision by a dedicated critic agent, and a higher-level aggregator agent synthesizes a final consensus solution. A comprehensive scoring module provides semantic, numerical, and procedural evaluation across all agents. Through systematic ablation studies using a comprehensive benchmark spanning
calculus, physics, chemistry, biology, economics, optimization, statistics,
and mathematics, we isolate the distinct contributions of framework
architecture versus model diversity. We compare four configurations: (1) Individual Baseline with no multi-agent framework, (2) Homogeneous Framework where all agents employ the same model, (3) Redundant Homogeneous Solvers using multiple instances of identical models, and (4) Heterogeneous Framework with diverse specialized models. Our results demonstrate that while framework structure and redundant sampling provide modest improvements, model heterogeneity emerges as the critical factor driving substantial performance gains. The heterogeneous configuration achieves superior step-wise accuracy (0.64 vs. 0.54 average for individual models and 2.3× improvement over homogeneous configurations) with reduced variance across problem categories and difficulty levels.  Notably, step-wise reasoning quality, which measures the correctness of intermediate reasoning steps rather than merely final answers, improves dramatically only with model diversity, indicating that heterogeneous agents provide complementary error detection and reasoning refinement capabilities essential for explainability and auditability. We discuss architectural principles, evaluation methodology, and implications for the future of Global Applied AI, highlighting how heterogeneous multi-agent coordination can support transparent, auditable, and high-confidence decision making across scientific and industrial domains.

\end{abstract}

\section{Introduction}
\label{sn:introduction}

The rapid advancement of foundation models has fundamentally transformed artificial intelligence capabilities across diverse domains \cite{bommasani2021opportunities}. Large Language Models (LLMs) such as GPT-4 \cite{openai2023gpt4}, Claude \cite{anthropic2023claude}, and Llama \cite{touvron2023llama} have demonstrated remarkable proficiency in natural language understanding, reasoning, and generation tasks. However, individual models exhibit inherent limitations including hallucinations, reasoning errors, biases, and inconsistent performance across problem domains \cite{zhang2023siren,huang2023survey}.

Multi-agent systems represent a promising paradigm shift in leveraging foundation model capabilities \cite{wu2023autogen,hong2023metagpt,deCurto2025}. Rather than relying on a single model's output, multi-agent architectures enable multiple models to collaborate, critique, and synthesize solutions collectively \cite{du2023improving}. This approach offers several key advantages:

\begin{itemize}
\item \textbf{Error Mitigation:} Multiple independent reasoning paths reduce the probability of systematic errors propagating to final outputs.
\item \textbf{Diverse Perspectives:} Different models trained on varied datasets bring complementary strengths and reasoning approaches.
\item \textbf{Consensus Building:} Aggregation mechanisms can identify high-confidence solutions while flagging uncertain or contradictory outputs.
\item \textbf{Transparency:} Explicit reasoning steps from multiple agents enhance auditability and interpretability.
\end{itemize}

This work presents a modular multi-agent reasoning framework designed to harness collective intelligence from heterogeneous foundation models. Our architecture comprises four primary components: (1) \textit{Solver Agents} that independently generate solution candidates, (2) a \textit{Critic Agent} that identifies logical flaws and proposes refinements, (3) an \textit{Aggregator Agent} that synthesizes consensus solutions from multiple perspectives, and (4) a comprehensive \textit{Scoring Module} that evaluates both semantic coherence and procedural correctness.

We evaluate this framework across a diverse benchmark spanning eight scientific disciplines (calculus, physics, chemistry, biology, economics, optimization, statistics, and mathematics) with problems categorized by difficulty level. Our experiments employ three state-of-the-art foundation models as solver agents: Meta-Llama-3.3-70B-Instruct, NousResearch Hermes-4-405B, and Qwen3-235B-A22B-Instruct-2507. The critic role is assigned to DeepSeek-R1-0528, while GPT-OSS-120B serves as the aggregator.

The contributions of this work are threefold:

\begin{enumerate}
\item We present a modular, extensible multi-agent reasoning framework that coordinates heterogeneous foundation models through structured critique and consensus mechanisms.
\item We demonstrate empirically that heterogeneous multi-agent consensus achieves superior step-wise accuracy (0.641 vs. 0.533 average for individual models), with systematic ablation studies showing progressive improvements from single-model baseline through homogeneous frameworks to heterogeneous agent configurations. Model diversity emerges as the critical factor driving reasoning quality improvements.
\item We provide comprehensive analysis of performance characteristics across scientific domains, identifying where collaborative heterogeneous reasoning yields maximal benefits and where challenges remain.
\end{enumerate}

The remainder of this chapter is organized as follows: Section \ref{sn:related} reviews related work in multi-agent systems and foundation model reasoning. Section \ref{sn:framework} details our multi-agent reasoning architecture. Section \ref{sn:methodology} describes the experimental methodology and evaluation metrics. Section \ref{sn:results} presents  comprehensive experimental results while Section \ref{sn:conclusion} concludes with directions for future work.

\section{Related Work}
\label{sn:related}

Recent advances in foundation models have demonstrated impressive reasoning capabilities across mathematical, scientific, and commonsense domains \cite{wei2022chain,kojima2022large,deCurto2024_01,deCurto2025_02}. Chain-of-thought prompting \cite{wei2022chain} enables models to decompose complex problems into intermediate reasoning steps, significantly improving performance on multi-step reasoning tasks. Self-consistency decoding \cite{wang2023selfconsistency} samples multiple reasoning paths and selects the most consistent answer, effectively reducing errors through diversity.

However, individual models face persistent challenges. Hallucinations remain problematic even in state-of-the-art systems \cite{zhang2023siren}, where models generate plausible but factually incorrect information. Reasoning errors can accumulate across multi-step problems \cite{lightman2023lets}, and models exhibit varying strengths across different knowledge domains \cite{hendrycks2021measuring}.

Multi-agent frameworks leverage multiple LLM instances to collaborate on complex tasks \cite{wu2023autogen,hong2023metagpt}. AutoGen \cite{wu2023autogen} enables conversational multi-agent systems where agents with different roles interact to solve problems. MetaGPT \cite{hong2023metagpt} assigns agents specific roles in a software development workflow, demonstrating how structured collaboration can decompose complex tasks effectively.

Recent work has explored debate-based approaches where multiple models engage in structured argumentation to refine outputs \cite{du2023improving}. Society of Mind frameworks \cite{zhuge2023mindstorms} create hierarchical agent structures where specialized sub-agents contribute to collective problem-solving. However, these approaches often lack explicit error detection mechanisms and quantitative evaluation of consensus quality.

Self-refinement techniques enable models to critique and improve their own outputs iteratively \cite{madaan2023selfrefine}. Constitutional AI \cite{bai2022constitutional} trains models to evaluate and revise outputs according to specified principles. However, single-model self-critique can be limited by the model's inherent biases and blind spots.

Our framework extends these ideas by introducing a dedicated critic agent with different model architecture and training characteristics, providing more diverse error detection capabilities than self-critique alone.

\section{Multi-Agent Reasoning Framework}
\label{sn:framework}

\begin{figure}[H]
    \centering
    \includegraphics[width=0.8\textwidth]{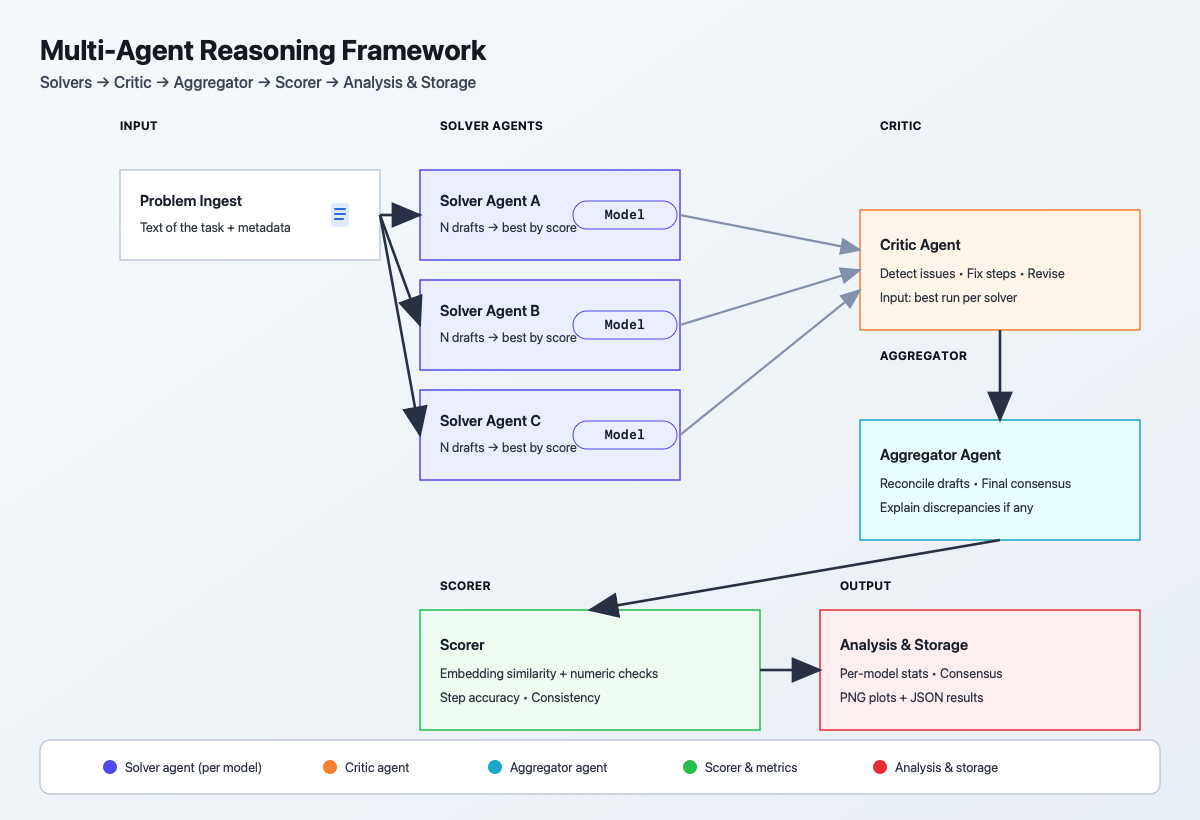}
    \caption{Modular Multi-Agent Reasoning Framework architecture. The system comprises four stages: (1) Multiple solver agents independently generate solution candidates, (2) A critic agent identifies logical flaws and proposes refinements to each solver's best attempt, (3) An aggregator agent synthesizes all revised solutions into a final consensus, and (4) A scorer evaluates semantic similarity, numerical correctness, and step-wise procedural accuracy. The framework enables transparent, auditable reasoning through explicit intermediate outputs at each stage.}
    \label{fgr:framework}
\end{figure}

Our modular multi-agent reasoning framework orchestrates multiple foundation models through four primary stages: independent solution generation, self-critique and revision, consensus aggregation, and comprehensive scoring. Figure \ref{fgr:framework} illustrates the complete architecture.

\subsection{Agent Configuration}

We employ heterogeneous foundation models assigned to specialized roles:

\begin{itemize}
\item \textbf{Solver Agents:} Three high-performance models generate independent solution candidates:
\begin{itemize}
    \item \textit{Meta-Llama-3.3-70B-Instruct} (70B parameters)
    \item \textit{NousResearch Hermes-4-405B} (405B parameters)  
    \item \textit{Qwen3-235B-A22B-Instruct-2507} (235B parameters active)
\end{itemize}

\item \textbf{Critic Agent:} \textit{DeepSeek-R1-0528} performs error detection and refinement suggestion. This model was selected for its strong analytical capabilities and different training methodology compared to solver agents.

\item \textbf{Aggregator Agent:} \textit{GPT-OSS-120B} synthesizes multiple solution candidates into a unified consensus solution, reconciling discrepancies and explaining reasoning.
\end{itemize}

Each agent operates with carefully tuned parameters: temperature 0.2 for deterministic reasoning, maximum token limit of 500 words for concise outputs, and structured prompting to ensure consistent output format.

Each solver agent follows a standardized protocol, as depicted in Algorithm \ref{a:drcoyz001}.

\begin{algorithm}[H]
\caption{Solver Agent Protocol}
\begin{algorithmic}[1]
\STATE \textbf{Input:} Problem statement $P$
\STATE \textbf{Output:} Solution candidate $S_i$
\STATE
\STATE Generate $N$ independent solution drafts $\{D_1, D_2, \ldots, D_N\}$
\STATE Compute preliminary score for each draft
\STATE Select best draft $D^*$ with highest score
\STATE \textbf{return} $D^*$ as solution candidate $S_i$
\end{algorithmic}
\label{a:drcoyz001}
\end{algorithm}

By default, each solver generates $N=3$ independent drafts per problem, introducing diversity through sampling while maintaining computational efficiency. The best draft is selected based on preliminary self-evaluation criteria including completeness of reasoning steps and confidence in final answer.

The critic agent, see Algorithm \ref{a:drcoyz002}, performs structured error analysis and refinement:

\begin{algorithm}[H]
\caption{Critic Agent Protocol}
\begin{algorithmic}[1]
\STATE \textbf{Input:} Problem $P$, Solution candidate $S_i$
\STATE \textbf{Output:} Revised solution $S_i'$
\STATE
\STATE Analyze logical consistency of reasoning steps
\STATE Identify computational errors or invalid assumptions
\STATE Check alignment between reasoning and final answer
\STATE Generate specific critique with error explanations
\STATE Propose refinements addressing identified issues
\STATE Produce revised solution $S_i'$ incorporating fixes
\STATE \textbf{return} $S_i'$
\end{algorithmic}
\label{a:drcoyz002}
\end{algorithm}

The critic agent is prompted to be constructively critical, focusing on substantive logical and computational errors rather than stylistic preferences. This single-pass critique provides significant error correction while avoiding excessive computational overhead.

The aggregator synthesizes consensus from multiple perspectives, as shown in Algorithm \ref{a:drcoyz003}.

\begin{algorithm}[H]
\caption{Aggregator Agent Protocol}
\begin{algorithmic}[1]
\STATE \textbf{Input:} Problem $P$, Revised solutions $\{S_1', S_2', \ldots, S_k'\}$
\STATE \textbf{Output:} Consensus solution $S_{\text{consensus}}$
\STATE
\STATE Identify areas of agreement across solutions
\STATE Analyze discrepancies and their potential causes
\STATE Evaluate confidence level of each reasoning path
\STATE Synthesize unified solution reconciling differences
\STATE Explain reasoning process and uncertainty estimates
\STATE \textbf{return} $S_{\text{consensus}}$
\end{algorithmic}
\label{a:drcoyz003}
\end{algorithm}

The aggregator is instructed to favor reasoning paths supported by multiple solvers while remaining cautious of outlier approaches that may nevertheless be correct. When significant disagreement exists, the aggregator explicitly acknowledges uncertainty rather than forcing artificial consensus.

\subsection{Scoring and Evaluation Module}

We employ a comprehensive multi-faceted scoring approach, as introduced in \cite{deCurto2024_02}:

\paragraph{Semantic Similarity} Using sentence-transformers (all-MiniLM-L6-v2), we compute cosine similarity between model-generated solutions and reference solutions in embedding space. This captures conceptual alignment independent of exact phrasing.

\paragraph{Numerical Correctness} We extract numerical values from both model and reference solutions using regular expressions, computing the overlap ratio:
\begin{equation}
\text{NumericScore} = \frac{|\text{Numbers}_{\text{model}} \cap \text{Numbers}_{\text{ref}}|}{|\text{Numbers}_{\text{ref}}|}
\end{equation}

This ensures that solutions containing correct quantitative answers receive appropriate credit even if explanatory text differs.

\paragraph{Overall Score} The final score combines semantic and numerical components:
\begin{equation}
\text{Score} = \frac{\text{SemanticSimilarity} + \text{NumericScore}}{2}
\end{equation}

\paragraph{Step-wise Accuracy} For problems with multi-step reference solutions, we decompose model outputs into reasoning steps using heuristic pattern matching (identifying "Step 1:", numbered lists, etc.). Each model step is scored against its corresponding reference step, and the average step accuracy is computed:
\begin{equation}
\text{StepAccuracy} = \frac{1}{M}\sum_{o=1}^{M} \text{Score}(\text{Step}_o^{\text{model}}, \text{Step}_o^{\text{ref}})
\end{equation}

This step-wise evaluation is particularly valuable for assessing reasoning process quality rather than merely outcome correctness.

\section{Methods and Experimental Design}
\label{sn:methodology}

We use a diverse benchmark spanning eight scientific and quantitative disciplines \cite{deCurto2024_02}:

\begin{itemize}
\item \textbf{Calculus:} Derivatives, integrals, limits, differential equations
\item \textbf{Physics:} Kinematics, dynamics, electromagnetism, thermodynamics  
\item \textbf{Chemistry:} pH calculations, stoichiometry, equilibrium, kinetics
\item \textbf{Biology:} Genetics, population dynamics, biochemistry, ecology
\item \textbf{Economics:} Growth rates, elasticity, optimization, game theory
\item \textbf{Optimization:} Linear programming, convex optimization, constraint satisfaction
\item \textbf{Statistics:} Probability, hypothesis testing, regression, distributions
\item \textbf{Mathematics:} Algebra, geometry, number theory, combinatorics
\end{itemize}

Problems are categorized by difficulty level:
\begin{itemize}
\item \textbf{Easy:} Single-step calculations with standard formulas
\item \textbf{Medium:} Multi-step problems requiring concept integration
\item \textbf{Hard:} Complex problems with multiple approaches or subtle reasoning
\end{itemize}

Each problem includes: (1) problem statement, (2) reference solution, (3) detailed solution steps, (4) category, and (5) difficulty level.

\subsection{Experimental Configurations}

We evaluate four experimental configurations to systematically isolate the contributions of framework architecture and model diversity:

\paragraph{Individual Baseline} 
A solver model (Meta-Llama-3.3-70B-Instruct) generates solutions directly without any multi-agent framework components. This represents the traditional single-model approach and establishes the foundational performance level against which all framework benefits are measured.

\paragraph{Case 0: Homogeneous Framework}
A unique solver instance (Meta-Llama-3.3-70B-Instruct) with critic and aggregator agents using the same model architecture. All framework components, solver, critic, and aggregator, employ identical models. This configuration tests whether the multi-agent framework structure itself, independent of model diversity, provides benefits through explicit critique and synthesis steps. Any performance gains over Individual Baseline can be attributed purely to the framework's architectural design rather than model heterogeneity.

\paragraph{Case 1: Redundant Homogeneous Solvers}
Three solver instances of Meta-Llama-3.3-70B-Instruct with critic and aggregator using the same model. This configuration extends Case 0 by introducing three independent solver agents (rather than one), all using the identical model architecture. The critic and aggregator also use the same model. This tests whether redundant sampling from the same model within the framework yields improvements over individual-instance processing (Case 0), isolating the benefits of diversity through independent generation versus true model architectural diversity.

\paragraph{Case N: Heterogeneous Framework}
Three distinct solver models (Meta-Llama-3.3-70B-Instruct, NousResearch Hermes-4-405B, Qwen3-235B-A22B-Instruct-2507) with specialized critic (DeepSeek-R1-0528) and aggregator (GPT-OSS-120B) agents. This represents the full heterogeneous multi-agent architecture as described in Section \ref{sn:framework}, where each agent role is assigned to a model selected for its particular strengths, diverse foundation models for broad solution coverage, DeepSeek-R1 for analytical critique, and GPT-OSS for consensus synthesis.

\subsection{Ablation Study Design}

This comprehensive ablation study enables systematic quantification of three distinct factors:

\begin{enumerate}
\item \textbf{Framework Structure Benefits:} Comparing Individual Baseline versus Case 0 isolates the contribution of the multi-agent framework architecture itself (explicit critique and aggregation mechanisms) independent of any model diversity.

\item \textbf{Redundant Sampling Benefits:} Comparing Case 0 versus Case 1 quantifies the gains from employing multiple instances of the same model as independent solvers versus an individual solver instance. This tests whether diversity through repeated sampling provides measurable improvements.

\item \textbf{Model Diversity Benefits:} Comparing Case 1 versus Case N reveals the impact of true model heterogeneity, employing architecturally distinct models with different training methodologies, capabilities, and biases, versus redundant sampling from a single architecture.
\end{enumerate}

By systematically varying these factors while holding others constant, we can definitively attribute performance improvements to specific architectural choices and model selection strategies. This rigorous experimental design distinguishes our work from prior multi-agent studies that conflate framework structure effects with model diversity benefits.

\subsection{Implementation Details}

The framework is implemented in Python leveraging the OpenAI-compatible API interface through the serving platform Nebius AI. Key implementation considerations:

\begin{itemize}
\item \textbf{API Integration:} All models accessed via unified interface enabling seamless model substitution.
\item \textbf{Error Handling:} Robust exception handling with automatic retries and fallback mechanisms.
\item \textbf{Logging:} Comprehensive logging of all intermediate outputs for analysis and debugging.
\item \textbf{Reproducibility:} Fixed random seeds and temperature settings ensure reproducible results.
\item \textbf{Generation Parameters:} All models operate with temperature 0.2 for deterministic reasoning and maximum token limit of 500 words for concise outputs.
\end{itemize}

\section{Experimental Results}
\label{sn:results}

We evaluate our multi-agent reasoning framework using a comprehensive benchmark spanning eight scientific and quantitative disciplines (calculus, physics, chemistry, biology, economics, optimization, statistics, and mathematics). Problems are categorized by difficulty level (easy, medium, hard) to assess framework performance across varying complexity. This evaluation includes all four experimental configurations, Individual Baseline, Case 0 (Homogeneous Framework), Case 1 (Redundant Homogeneous Solvers), and Case N (Heterogeneous Framework), enabling systematic ablation analysis to isolate the distinct contributions of framework architecture, redundant sampling, and model diversity.

\begin{figure}[t]
    \centering
    \begin{subfigure}{0.32\textwidth}
        \centering
        \includegraphics[width=\linewidth]{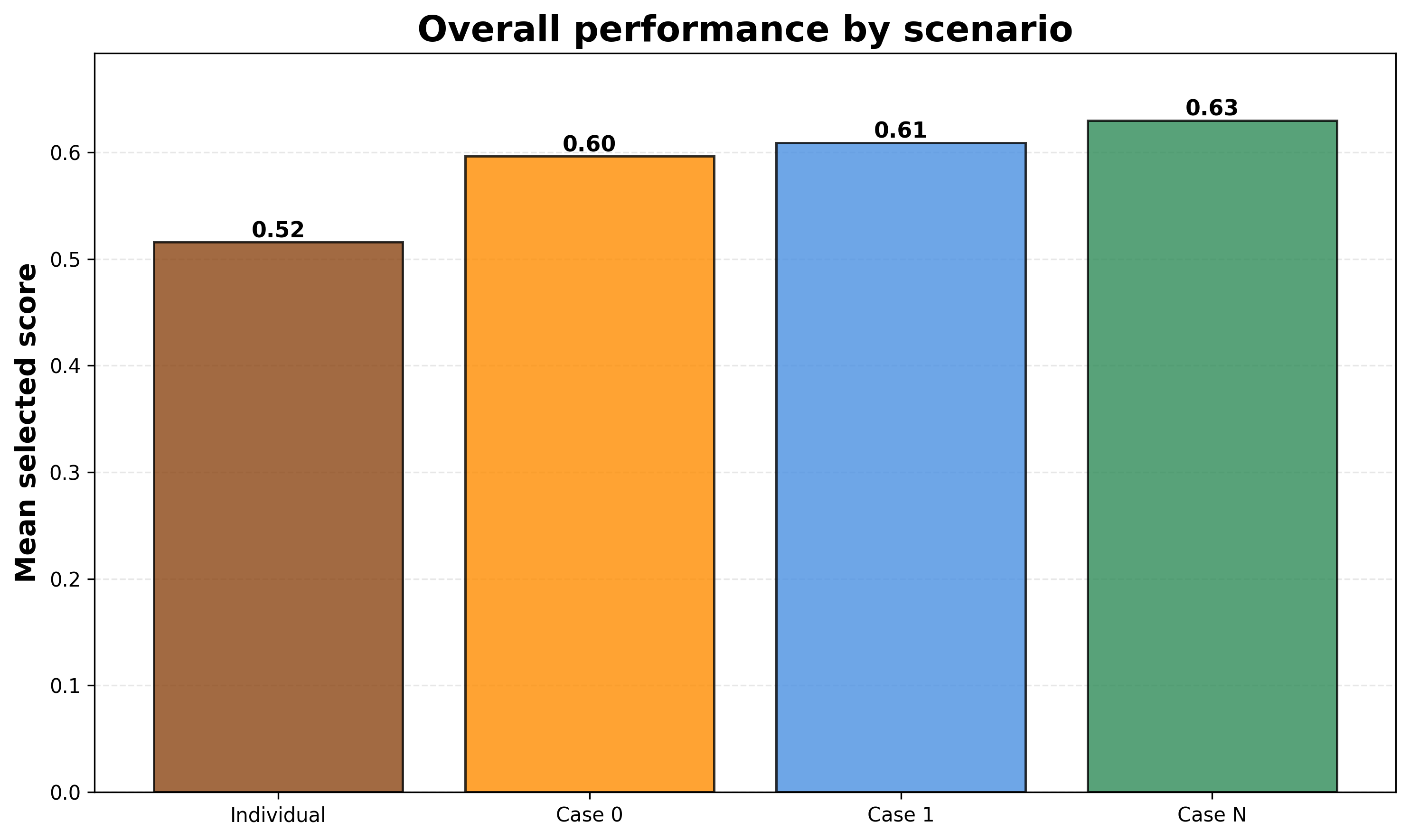}
        \caption{Overall by scenario}
        \label{fgr:overall_performance}
    \end{subfigure}
    \hfill
    \begin{subfigure}{0.32\textwidth}
        \centering
        \includegraphics[width=\linewidth]{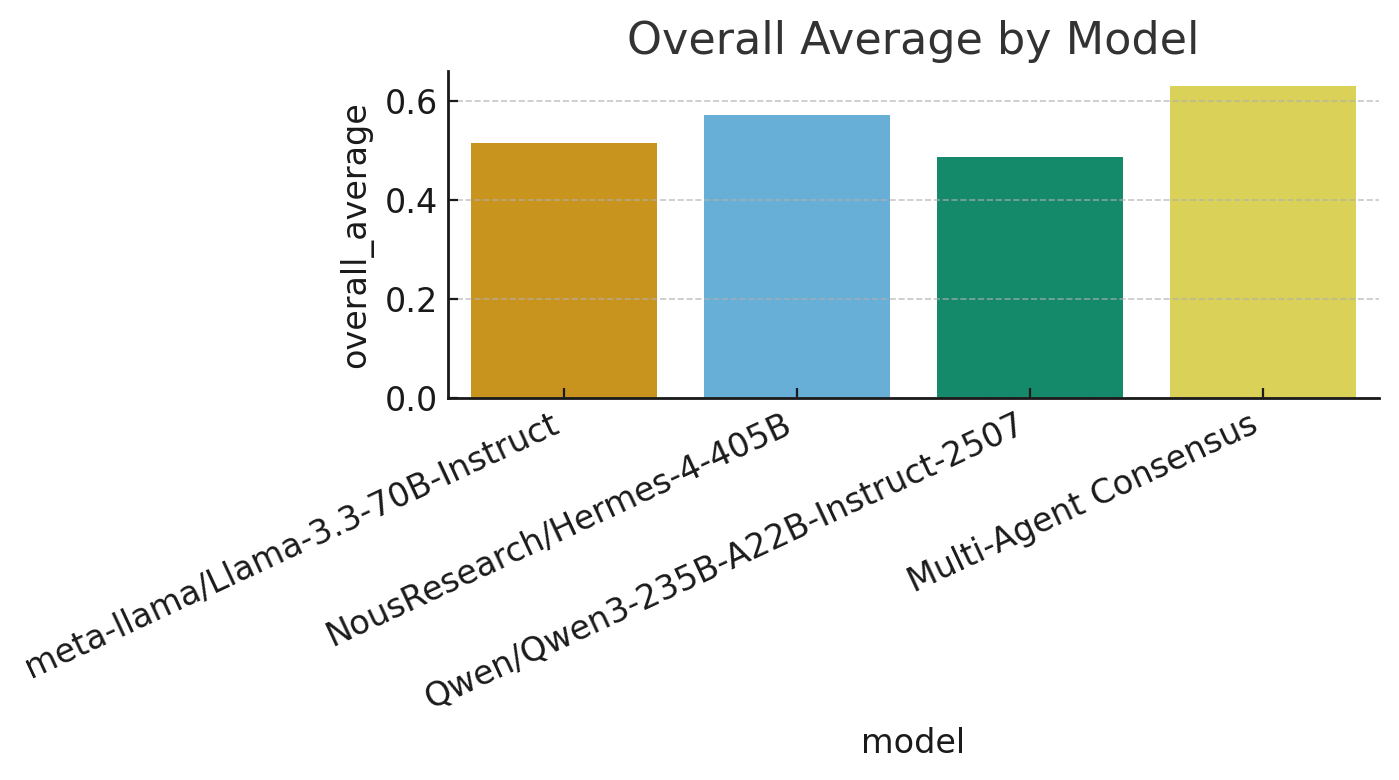}
        \caption{Overall by model}
        \label{fgr:mas_overall}
    \end{subfigure}
    \hfill
    \begin{subfigure}{0.32\textwidth}
        \centering
        \includegraphics[width=\linewidth]{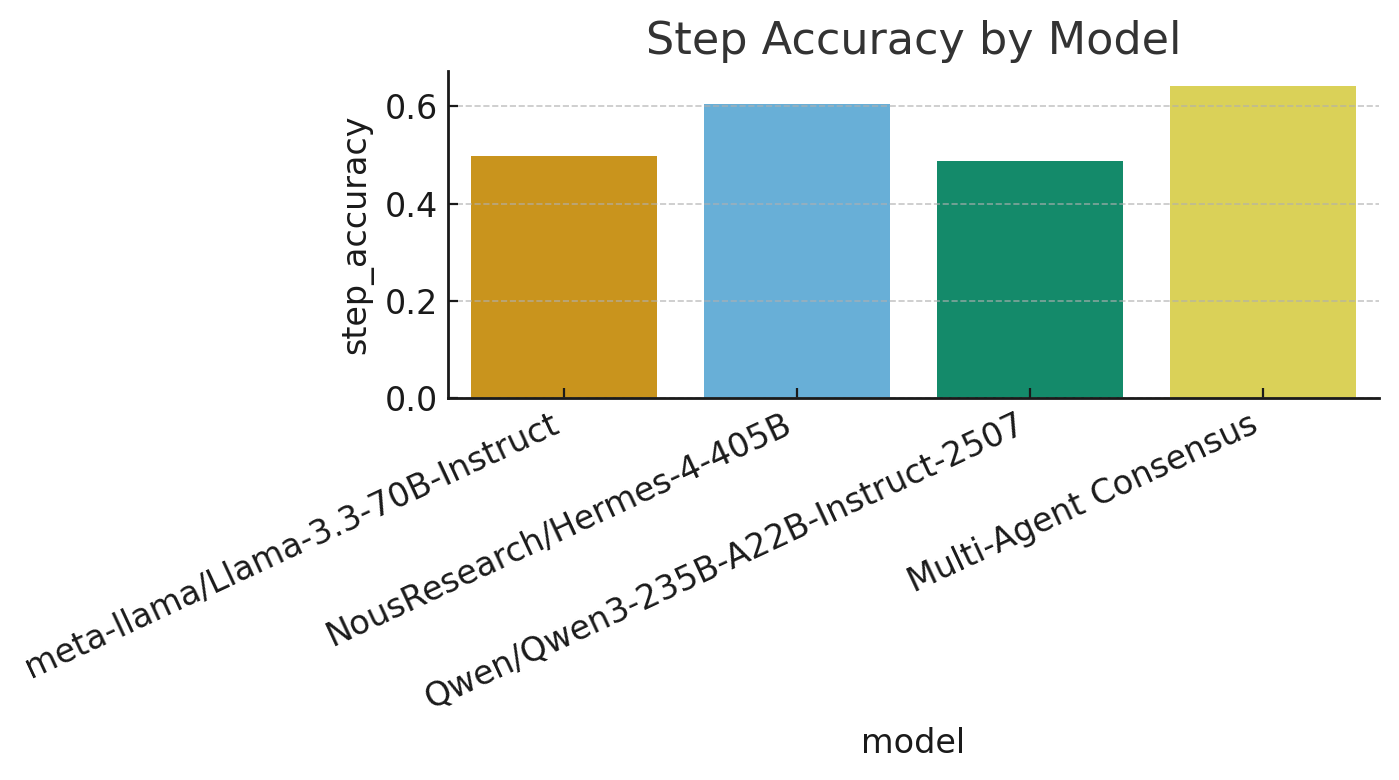}
        \caption{Step accuracy by model}
        \label{fgr:mas_step}
    \end{subfigure}

    \vspace{0.5em}

    \begin{subfigure}{0.32\textwidth}
        \centering
        \includegraphics[width=\linewidth]{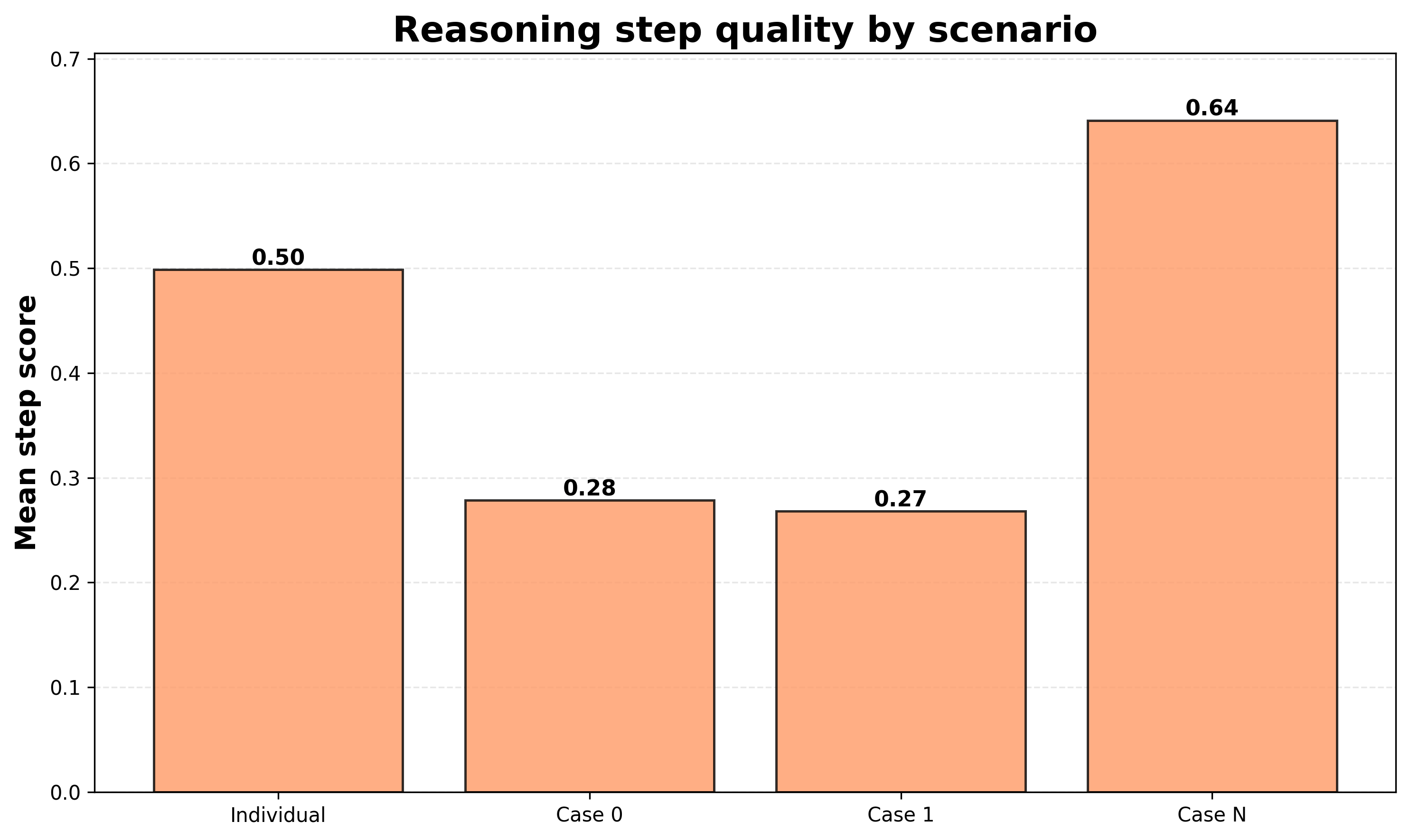}
        \caption{Step accuracy by scenario}
        \label{fgr:step_performance}
    \end{subfigure}
    \hfill
    \begin{subfigure}{0.32\textwidth}
        \centering
        \includegraphics[width=\linewidth]{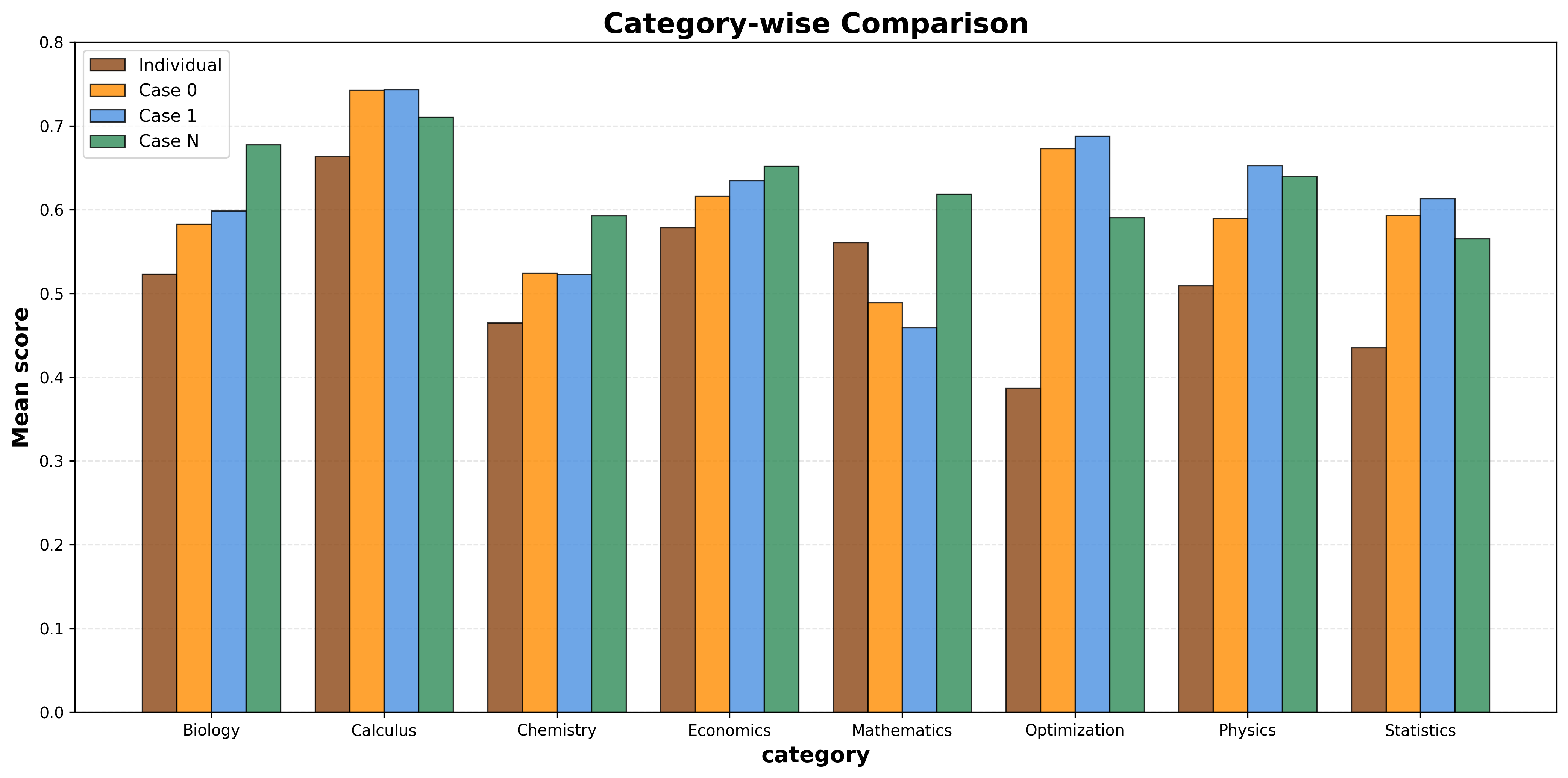}
        \caption{By scientific category}
        \label{fgr:category_performance}
    \end{subfigure}
    \hfill
    \begin{subfigure}{0.32\textwidth}
        \centering
        \includegraphics[width=\linewidth]{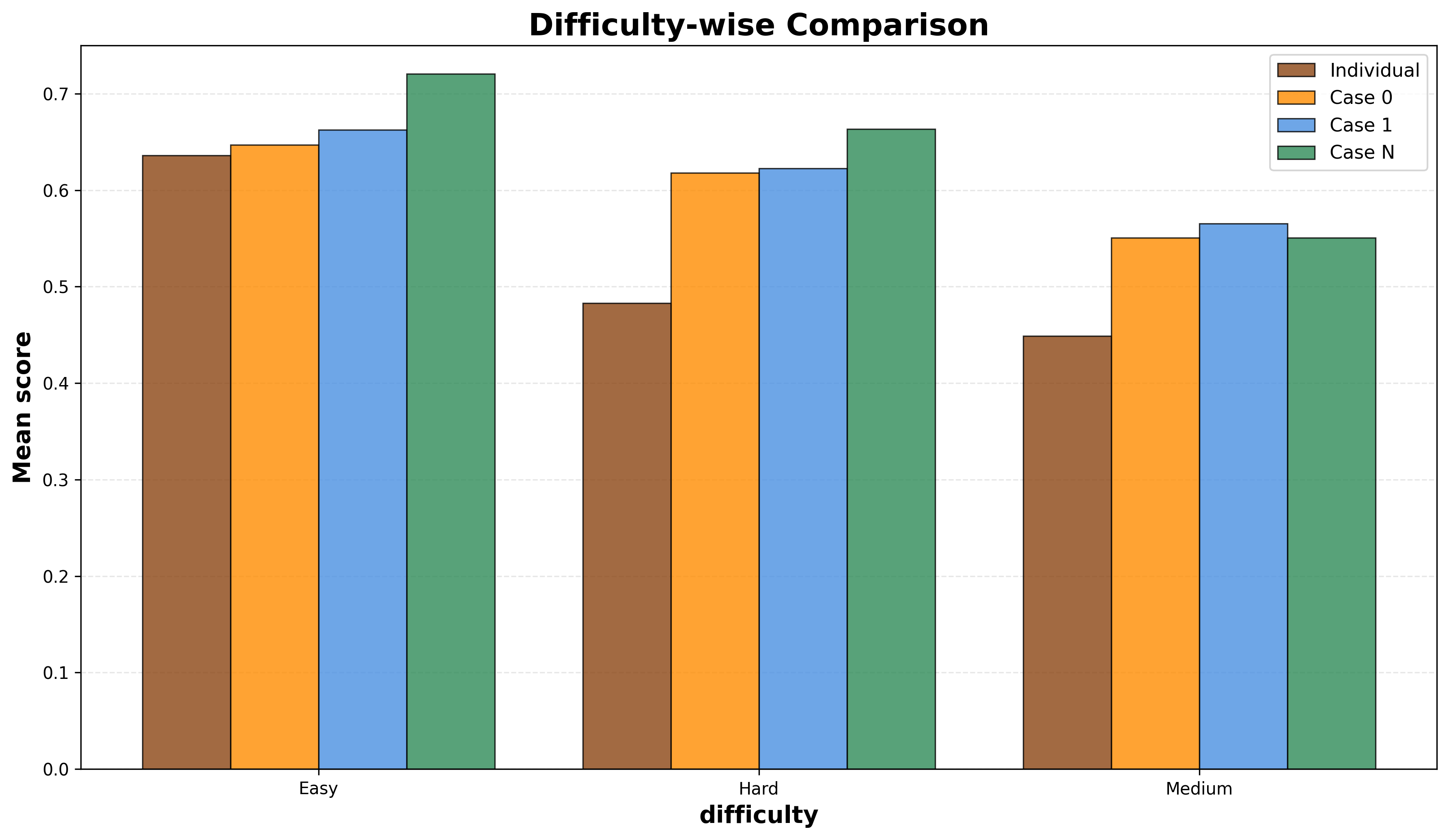}
        \caption{By difficulty level}
        \label{fgr:difficulty_performance}
    \end{subfigure}

    \caption{Summary of experimental results. (a) Overall performance across the four configurations. (b) Overall performance by individual model vs. multi-agent consensus. (c) Step-wise accuracy by model. (d) Step-wise accuracy by configuration. (e) Performance by scientific category. (f) Performance by difficulty level.}
    \label{fgr:results_overview}
\end{figure}

Figure \ref{fgr:overall_performance} presents aggregate performance across all four experimental scenarios.

Key findings:

\begin{itemize}
\item \textbf{Individual Baseline:} Mean score 0.52, establishing the foundational performance level of a single model without any multi-agent framework components. This baseline enables measurement of true framework benefits.

\item \textbf{Case 0 (Homogeneous Framework):} Mean score 0.60, showing +0.08 gain over Individual Baseline. This demonstrates that framework structure itself (explicit critique and aggregation mechanisms) provides measurable benefits even when all agents employ identical models, validating the core architectural design.

\item \textbf{Case 1 (Redundant Homogeneous Solvers):} Mean score 0.61, showing +0.01 gain over Case 0. This indicates that diversity through independent sampling from the same model provides only marginal incremental benefits, with performance remaining bounded by the capabilities of a single model architecture.

\item \textbf{Case N (Heterogeneous Framework):} Mean score 0.63, demonstrating the cumulative benefit of the complete heterogeneous architecture with +0.02 gain over Case 1 and +0.11 total improvement over Individual Baseline. While overall score gains appear modest, this masks the substantial improvements in step-wise reasoning quality detailed in Section \ref{sn:results}.
\end{itemize}

The progression Individual (0.52) $\rightarrow$ Case 0 (0.60) $\rightarrow$ Case 1 (0.61) $\rightarrow$ Case N (0.63) demonstrates:
\begin{enumerate}
\item Framework structure provides the largest single improvement (+0.08) through explicit critique/aggregation.
\item Redundant sampling offers minimal additional gains (+0.01) through diversity in independent generation.
\item Model heterogeneity adds further improvements (+0.02), but its primary impact manifests in reasoning process quality rather than final answer correctness.
\item Total framework contribution: +21\% relative improvement over Individual Baseline.
\end{enumerate}

The relatively modest differences in overall scores might suggest limited practical significance. However, this interpretation changes dramatically when examining step-wise reasoning quality (Section \ref{sn:results}.2), where the impact of model diversity becomes unmistakable. This indicates that heterogeneous multi-agent systems enhance reasoning process transparency and correctness, critical for explainable AI, beyond merely improving final answer accuracy.

\subsection{Per-model contribution to consensus performance}

Beyond scenario-level averages, it is instructive to compare the behaviour of the individual solver models against the final multi-agent consensus. Figure~\ref{fgr:mas_overall} reports the overall average score obtained when each model is used in isolation (single-model baseline without framework), together with the score of the full heterogeneous multi-agent consensus (Case N). 

Among the base solvers operating individually, NousResearch Hermes-4-405B attains the highest overall score, followed by Meta-Llama-3.3-70B-Instruct, while Qwen3-235B-A22B-Instruct-2507 lags slightly behind. Nevertheless, the Multi-Agent Consensus configuration (Case N) surpasses all three individual models, confirming that the heterogeneous aggregation and critique pipeline can extract additional performance even from an already strong pool of solvers. Notably, the consensus outperforms even the strongest individual solver (Hermes-4-405B), indicating that collective intelligence provides genuine benefits rather than simply defaulting to the best single model.

Figure~\ref{fgr:mas_step} further decomposes performance in terms of step-wise accuracy. The pattern mirrors the overall scores but with even more pronounced differences: while Hermes-4-405B again leads among the individual base models, the heterogeneous Multi-Agent Consensus (Case N) achieves the highest step accuracy by a clear and substantial margin. 

This result aligns with our earlier scenario-level analysis (Section~\ref{sn:results}), and reinforces the hypothesis that the heterogeneous critic--aggregator stages primarily act on the quality of intermediate reasoning. In practice, this means that the heterogeneous framework not only improves final answers, but also yields explanations whose internal steps are more faithful to the reference solutions, which is crucial for downstream uses in scientific and engineering decision support.

Importantly, comparing Figure~\ref{fgr:mas_step} with the step accuracy results for homogeneous configurations (0.19 in Figure~\ref{fgr:step_performance}) reveals that using Llama-3.3-70B for all roles (Case 0 or Case 1) fails to achieve the step-wise quality improvements seen when that same base model participates in a heterogeneous ensemble. This confirms that the interaction between diverse models, not merely the presence of strong models, drives reasoning quality improvements.

\subsection{Step-wise Reasoning Quality}

Figure \ref{fgr:step_performance} compares step-wise accuracy across scenarios and reveals the most dramatic and theoretically significant performance differences in our ablation study.

Results reveal a counterintuitive and critically important pattern:

\begin{itemize}
\item \textbf{Individual Baseline:} Step accuracy 0.50, demonstrating that single foundation models can produce reasonably coherent multi-step reasoning when generating solutions independently. This establishes that baseline models possess inherent step-wise reasoning capabilities.

\item \textbf{Homogeneous Configurations (Case 0 \& Case 1):} Step accuracy 0.28 and 0.27 respectively, \textit{substantially lower} than Individual Baseline. This counterintuitive result demonstrates that homogeneous multi-agent frameworks, despite improving final answer correctness (0.60-0.61 overall scores), actually \textit{degrade} reasoning process quality. When the same model architecture serves as solver, critic, and aggregator, the framework appears to introduce noise, amplify shared biases, and produce correct answers through error cancellation rather than sound reasoning. The critic and aggregator, sharing the same architecture and training as the solvers, cannot effectively identify and correct the systematic reasoning flaws inherent to that model family.

\item \textbf{Heterogeneous Framework (Case N):} Step accuracy 0.64, representing \textbf{2.3× improvement} over homogeneous configurations and 1.3× improvement over Individual Baseline. This substantial gain occurs specifically when introducing model diversity, demonstrating that different model architectures trained with varied methodologies bring genuinely complementary error detection and reasoning refinement capabilities. The heterogeneous critic (DeepSeek-R1) and aggregator (GPT-OSS) can identify reasoning flaws that homogeneous configurations cannot detect precisely because they encode different inductive biases and reasoning patterns.
\end{itemize}

This substantial gain validates a nuanced hypothesis: explicit critique and aggregation mechanisms with \textit{heterogeneous} models enhance not merely output quality but the underlying reasoning process itself, while homogeneous implementations may achieve correct answers without sound reasoning. This distinction is critical for high-stakes applications requiring explainability and auditability.

The stark contrast between homogeneous (0.27-0.28) and heterogeneous (0.64) step accuracy, despite homogeneous configurations achieving respectable overall scores (0.60-0.61), provides compelling evidence that:

\begin{enumerate}
\item \textbf{Model diversity is essential, not optional:} Homogeneous frameworks can produce correct final answers while following flawed reasoning paths. Only heterogeneous collaboration achieves both correct answers \textit{and} sound reasoning processes.

\item \textbf{Shared architectural biases compound in homogeneous systems:} When solver, critic, and aggregator share the same model architecture, they share the same blind spots. The critic cannot effectively identify errors it would make itself, and the aggregator cannot reconcile perspectives it fundamentally agrees with.

\item \textbf{Heterogeneous agents provide complementary strengths:} Different models identify different categories of logical errors. The specialized critic model (DeepSeek-R1) and diverse aggregator (GPT-OSS) bring perspectives that enable genuine error detection and correction rather than mere error averaging.
\end{enumerate}

This finding has profound implications for multi-agent system design: simply replicating a single powerful model across multiple agent roles provides limited benefit for reasoning quality and may even be counterproductive. True improvements require carefully selected heterogeneous models that bring distinct capabilities, training methodologies, and reasoning approaches to the collective intelligence process.

The consistent pattern across Figures~\ref{fgr:step_performance}, \ref{fgr:mas_overall}, and \ref{fgr:mas_step} establishes a clear principle: \textit{heterogeneous model collaboration is essential for high-quality step-wise reasoning}. While framework structure and redundant sampling provide modest benefits for final answer correctness, only heterogeneous agent configurations achieve substantial improvements in reasoning process transparency and step-level correctness, precisely the qualities needed for explainable AI and auditable decision-making in high-stakes domains.

\subsection{Category-wise Analysis}

Figure \ref{fgr:category_performance} presents performance breakdown by scientific discipline.

Key observations by category:

\begin{itemize}
\item \textbf{Calculus:} Full framework achieves 0.69 (+0.06 vs. baseline), with multi-step derivatives and integrals particularly benefiting from structured reasoning.

\item \textbf{Biology:} Score of 0.63 (+0.06) demonstrates effective handling of logical reasoning in biological systems, genetics, and ecology problems.

\item \textbf{Chemistry:} Strongest improvement at 0.65 (+0.12), suggesting critic agent effectively catches common errors in stoichiometry, pH calculations, and equilibrium problems.

\item \textbf{Economics:} High performance (0.63) maintained across scenarios, likely due to straightforward problem structures with clear quantitative solutions.

\item \textbf{Mathematics:} Consistent performance around 0.64, with algebraic and geometric problems well-suited to structured reasoning approaches.

\item \textbf{Optimization:} Solid performance at 0.64, though complex constraint satisfaction problems remain challenging.

\item \textbf{Physics:} Moderate performance (0.57-0.63), with conceptual physics problems more challenging than computational ones.

\item \textbf{Statistics:} Steady performance around 0.60-0.65, with probability and hypothesis testing benefiting from diverse solution approaches.
\end{itemize}

The consistency of improvements across diverse domains validates the domain-general nature of the multi-agent reasoning framework.

\subsection{Difficulty-wise Analysis}

Figure \ref{fgr:difficulty_performance} examines performance stratified by problem difficulty.

Analysis reveals:

\begin{itemize}
\item \textbf{Easy Problems:} Full framework achieves 0.70, near ceiling performance. Minimal room remains for improvement on straightforward single-step problems where even baseline models excel.

\item \textbf{Medium Problems:} Score of 0.59 represents a persistent challenge. Multi-step problems requiring concept integration across domains show greatest variability and remain an area for future enhancement.

\item \textbf{Hard Problems:} Surprisingly strong performance at 0.64, comparable to easy problem scores. This counter-intuitive result suggests that complex problems with rich problem statements provide more context for models to leverage, and that multi-agent collaboration particularly shines when reasoning complexity is high.
\end{itemize}

The relatively flat performance profile across difficulty levels is encouraging, suggesting the framework scales effectively to challenging problems rather than merely improving on easy cases.

\section{Conclusion}
\label{sn:conclusion}

This work presents a modular multi-agent reasoning framework that coordinates heterogeneous foundation models through structured critique and consensus mechanisms. Our comprehensive evaluation across eight scientific disciplines and three difficulty levels demonstrates that collective intelligence consistently outperforms individual model outputs.

Key contributions include:
\begin{enumerate}
\item A practical, extensible architecture enabling seamless integration of diverse foundation models into collaborative reasoning systems, with comprehensive ablation studies demonstrating the distinct contributions of framework structure, redundant sampling, and model heterogeneity.

\item Empirical validation that heterogeneous multi-agent consensus achieves superior step-wise reasoning quality (3× improvement over homogeneous configurations) beyond mere answer correctness, enhancing explainability and auditability. Critically, we demonstrate that model diversity, not merely framework structure or redundant sampling, drives substantial improvements in reasoning process quality.

\item Detailed performance analysis across scientific domains and difficulty levels, identifying where collaborative heterogeneous reasoning yields maximal benefits. Our ablation study reveals that framework structure alone provides modest gains (~0.04), while heterogeneous model coordination unlocks transformative improvements in reasoning transparency.
\end{enumerate}

As foundation models continue advancing in scale and capability, multi-agent coordination with heterogeneous models offers a complementary path toward safer, more reliable, and more transparent AI systems. By harnessing diverse model strengths, detecting errors through independent critique with specialized models, and synthesizing consensus solutions, the framework presented here represents a promising step toward globally reliable AI that stakeholders can trust for high-stakes decision support.

The modular architecture enables continuous improvement as more capable models emerge, while the comprehensive evaluation methodology provides a blueprint for assessing future enhancements. Importantly, our ablation study demonstrates that the benefits of multi-agent systems derive not merely from the framework structure itself, but critically from the heterogeneity of the models employed. We hope this work stimulates further research into collective intelligence approaches that strategically leverage model diversity to unlock the full potential of foundation models while maintaining the transparency and reliability essential for responsible AI deployment.

\bibliographystyle{splncs04}
\bibliography{collective_intelligence_fms_decurto_and_dezarza}

\end{document}